# Migration Induced Epidemics: Dynamics of Flux-Based Multipatch Models

by Larry S. Liebovitch[1,2] and Ira B. Schwartz[1]*

[1]Naval Research Laboratory, Code 6792, Plasma Physics Division, Washington DC 20375
[2]Florida Atlantic University, Center for Complex Systems and Brain Sciences, Center for Molecular Biology and Biotechnology, Department of Psychology, Boca Raton FL 33431

**Abstract**

Classical disease models use a mass action term as the interaction between infected and susceptible people in separate patches. We derive the equations when this interaction is a migration of people between patches. The results model what happens when a new population is moved into a region with endemic disease.

*__Corresponding Author__
Ira B. Schwartz
Naval Research Laboratory
Code 6792
Plasma Physics Division
Washington DC 20375
Telephone: 202.404.8359
Fax: 202.767.0631
E-mail:schwartz@nlschaos.nrl.navy.mil

The spread of infectious disease in a population can be modeled by a set of ordinary differential equations that describe the rate at which the populations of susceptible, exposed, infected, and recovered depend on the number of people in those categories and a set of parameters that model the infectious spread and recovery of that disease [1,2]. In the simplest form of this model we deal only with the number of susceptible, S, and infected individuals I, namely:

$$\frac{dS}{dt} = \mu N - (\frac{\beta}{N})IS \qquad (1)$$

$$\frac{dI}{dt} = (\frac{\beta}{N})IS - \gamma I \qquad (2)$$

where $\mu$ is the birthrate at which new susceptibles are added to the population, N is the total number of people, ($\beta$/N) is the contact rate, and $\gamma$ is the rate at which infected people recover. Recovered individuals are assumed to be permanently immune and for populations of constant size may be neglected. A central assumption in these models is that the rate of new infections is proportional to the mass action term ($\beta$/N)IS.

These models assume that the infected and susceptible people are perfectly well mixed at every instant in time. An increasingly important question in epidemiology is how to extend these classical formulations to adequately describe the spatial heterogeneity in the distribution of susceptible and infected people and in the parameters of the spread of the infection that is observed in both experimental data and computer simulations [3-5]. This has been done previously by modeling the people as organized into separate patches of different sizes and assuming that the rate of new infections in patch k due to the infected people in patch j also has the form ($\beta_{kj}/N_k)I_jS_k$. That is, the coupling between the patches has been assumed to be a nonlinear, mass action term. Such coupling, along with seasonal driving, has been shown to excite long period, small amplitude oscillations in both deterministic and stochastic settings [6].

Here we propose a new approach to compute the effects of the spatial heterogeneity and the spread of infection between different patches of people. First, we describe how the parameters of the spread of infection are likely to scale with the number of people in a patch. This makes it possible to compute the number of susceptible and infected people at steady state and compare those results from sets of patches with different distributions of population sizes.

Second, instead of the classical mass action approach, we model the spread of infection between the patches as the migration, that is a flux, of susceptible and infected people between the patches. We show that this new flux-based approach is useful in computing the steady state and dynamical properties of the spread of infection through different patches. We use it to compute how the migration of people between the patches alters the steady state number of susceptible and infected people in a distribution of patches of different sizes. We show, perhaps surprisingly, that the migration of infected people changes the steady state number of infected people in each patch, but not the steady state total number of infected people in all the patches. We then use the flux-based approach to compute the dynamical behavior of patches when the onset of migration occurs at different time scales. These results may be useful in understanding what happens when a new population is introduced into a region with endemic disease. We show that epidemics result when the onset of migration is rapid compared to natural time scale of the patches. In models with two patches we determine how the maximum number of people infected in the epidemic depends on the onset time scale of the migration. In models with 20 patches, we show that this epidemic can appear to travel as a wave spreading through the patches.

Eqs. (1-2) are based on the assumption that the entire population is homogenous. In a real population, it is more likely that an infected person will spread the disease to nearby susceptibles, while further away there will remain homogenous pools of non-infected susceptibles. Thus, there will be islands of both susceptible and infected people. The spread of the disease will depend on the complex geometry of the sizes of these regions and the borders between them. This patchiness of susceptibles is analogous to diffusion limited chemical reactions in a tank reactor, such as A + B -> C [7]. As the A's and B's interact in a region they are converted to C's except for excess A's or B's. Thus, as time goes on, increasingly large regions of A's, B's, and C's are generated. The reaction now can only proceed on the borders of those regions. Thus the reaction rate declines in time, whereas it is normally assumed to be a constant in time. The reaction rate is constant only when, at each instant, an invisible hand reaches into the reaction components and well mixes all the reactants and products. The nature of the reaction itself creates spatial heterogeneities that change the dynamics of the reaction.

Because the regions around an infected person become depleted with susceptible people who have just become infected, the rate of infection will be less than if all of the susceptibles in the population were at risk. Thus $\beta$ will depend on the size of the population N in the patch. As N increases, the screening effect of the heterogeneity also increases. Thus $\beta$, which is a constant in the classical model, is more realistically a function of N. We assume the anzats that

$$\beta = \beta_s N^{-\delta} \qquad (3)$$

where $\beta_s$ is a constant, N is the number of people in a patch, and $|\delta|<1$. Such power law scalings have been described by Anderson and May [1]. Based on whether an infected person always contacts the same number of other people, or a number of people proportional to the population size, they proposed that $-1 < \delta < 0$ and estimated that -0.07 $< \delta < -0.03$. On the other hand, Hethcote et al. [8-11] found that for 5 human diseases $\delta = -0.05$. Because of the uncertainty in the value of $\delta$, for completeness, we will consider both the cases for $\delta \leq 0$ and $\delta > 0$. Similarly, we also assume that the birthrate also depends N, namely

$$\mu = \mu_s N^\varepsilon \qquad (4)$$

where $\mu_S$ is a constant and $|\varepsilon|<1$. We consider the cases for both $\varepsilon \leq 0$ and $\varepsilon > 0$, that is when the susceptibles are preferentially added further away from the infected people as well as closer to the infected people.

We first determine how the steady state total number of susceptible, $S_{total}$, and infected, $I_{total}$, people compare to those that would occur when are all $N_T$ are present in one homogeneous patch for different distribution of patch sizes. The steady state solutions, $dS/dt = dI/dt = 0$, to Eqs. (1-2) are $S = (\gamma/\beta) N$ and $I = (\mu/\gamma) N$. If the whole population, $N_T$, is divided into $(N_T/N_0)$ patches each of population $N_0$ then,

$$\frac{S_{total}}{S_w} = \left(\frac{N_0}{N_T}\right)^\delta \qquad (5)$$

$$\frac{I_{total}}{I_w} = \left(\frac{N_0}{N_T}\right)^\varepsilon \qquad (6)$$

where $S_w$ and $I_w$ are the total number of susceptible and infected people when the whole population is present in one patch.

Many physical, chemical, and biological systems with spatial heterogeneity are well characterized by power law distributions in the distribution of the sizes of the spatial domains [12-16]. Moreover, the underlying mechanisms that produce the spatial heterogeneity in those systems, such as those in diffusion limited chemical reactions, often depend on interactions that occur only at the borders of the spatial domains, which here is analogous to the spread of infection from infecteds to susceptibles. Therefore it is instructive to consider what happens when the population is partitioned into f(N) patches with N people in each patch, where

$$f(N) = A N^{-D} \qquad (7)$$

and the smallest patch has $N_0$ people and the largest patch has $\alpha N_0$ people, where $\alpha > 1$. The total number of people in all the patches, $N_T$ is

$$N_T = \int_{N_0}^{\alpha N_0} N\, f(N)\, dN, \tag{8}$$

and by integrating $I(N)f(N)dN$ and $S(N)f(N)dN$ we find that

$$\frac{S_{total}}{S_w} = \left(\frac{N_0}{N_T}\right)^{\delta} \left(\frac{2-D}{2-D+\delta}\right) \left(\frac{\alpha^{2-D+\delta}-1}{\alpha^{2-D}-1}\right) \tag{9}$$

$$\frac{I_{total}}{I_w} = \left(\frac{N_0}{N_T}\right)^{\varepsilon} \left(\frac{2-D}{2-D+\varepsilon}\right) \left(\frac{\alpha^{2-D+\varepsilon}-1}{\alpha^{2-D}-1}\right) \tag{10}$$

The total steady state number of susceptible and infected people in a heterogeneous set of patches, each with different infectious parameters $\beta$ and $\mu$, are given by Eqs. (5-6) and (9-10). For example, for a power law distribution of populations, when $\delta > 0$ and $\varepsilon > 0$, the total steady state number of infected and susceptible people is less than those that would be present if all the people were in a single, well mixed patch with $\beta$ and $\mu$ given by Eqs. (3-4). The difference from the single, well mixed patch increases as D increases in Eq. (7), and reaches plateaus as D approaches $\pm\infty$. For a power law distribution of populations, when $\delta < 0$ and $\varepsilon < 0$, then the total steady state number of infected and susceptible people is greater than those that would be present if all the people were in a single, well mixed patch and this difference increases with increasing D. The utility of this approach, based on the scalings of the infectious parameters with population size and on the distribution of patch size, is that it makes it possible to quantitatively determine how the steady number of susceptible and infected people depends on the heterogeneity of the patches.

So far we have modeled the spatial heterogeneity by partitioning the population into separate, non-interacting patches. We now determine how these results are changed when there is a migration of susceptible and infected people between the patches. Here we consider the physical movement of infected people from one patch to another, for example, such as the relocation of people from one city to another or in refugee camps [17]. Note that this flux-based approach differs from models which use a mass action term to model the interaction between the patches. This change in perspective makes it possible for us to show that when susceptible or infected people move from one patch into another that there is no change in the steady state total number infected and little change in the steady state total number of susceptibles.

We start by considering only one patch with a constant net rate of $p^s$ susceptible people and $p^i$ infected people moving into that patch. Using the fraction of susceptibles $s = S/N$, and the fraction of infecteds $i = I/N$, Eqs. (1-2) now become

$$\frac{ds}{dt} = \mu - \beta is + p^s \tag{11}$$

$$\frac{di}{dt} = \beta is - \gamma i + p^i \tag{12}$$

The steady state solutions, $ds/dt = di/dt = 0$, for Eqs. (11) and (12) are

$$s = \left(\frac{\gamma}{\beta}\right)\left[\frac{1 + \frac{p^s}{\mu}}{1 + \frac{(p^s + p^i)}{\mu}}\right] \tag{13}$$

$$i = \left(\frac{\mu}{\gamma}\right) + \left(\frac{p^s}{\gamma}\right) + \left(\frac{p^i}{\gamma}\right) \tag{14}$$

If there were no movement of susceptible or infected people, then $p^s = p^i = 0$, and the fraction of susceptibles $s^0 = (\gamma/\beta)$ and infecteds $i^0 = (\mu/\gamma)$. We now consider L patches with parameters $\beta_k$ and $\mu_k$ in each patch k, where there is no movement of people between the patches. The total number of susceptibles and infected is

$$S_{total}^0 = \sum_{k=1}^{L} \left(\frac{\gamma}{\beta_k}\right) N_k \tag{15}$$

$$I_{total}^0 = \sum_{k=1}^{L} \left(\frac{\mu_k}{\gamma}\right) N_k \tag{16}$$

When susceptible and infected people move between the L patches, the total number of infected in all the patches, found from Eq. (14) becomes

$$I_{total} = \sum_{k=1}^{L} \left(\frac{\mu_k}{\gamma}\right) N_k + \left(\frac{1}{\gamma}\right)\sum_{k=1}^{L} p_k^s N_k + \left(\frac{1}{\gamma}\right)\sum_{k=1}^{L} p_k^i N_k. \tag{17}$$

Since the all the susceptible and infected people that leave one patch must enter another patch, the net flow of susceptible and infected people must equal zero, namely,

$$\sum_{k=1}^{L} p_k^s N_k = \sum_{k=1}^{L} p_k^i N_k = 0 \tag{18}$$

and thus

$$I_{total} = \sum_{k=1}^{L} \left(\frac{\mu_k}{\gamma}\right) N_k = I_{total}^0 \qquad (19)$$

Hence, the movement of susceptible and infected people from one patch to another patch does not change the steady state total number of infected people. The movement changes the number of infected people at the steady state in each patch, but it does not change the total, steady state number of infected summed over all the patches.

The total number of susceptible in all the patches, found from Eq. (13) becomes

$$S_{total} = \sum_{k=1}^{L} \left(\frac{\gamma}{\beta_k}\right) \left[ \frac{1 + \frac{p_k^s}{\mu_k}}{1 + \frac{(p_k^s + p_k^i)}{\mu_k}} \right] N_k \qquad (20)$$

When $(p_k^s/\mu_k)$, $(p_k^i/\mu_k) \ll 1$, then the total number of susceptibles becomes

$$S_{total} = \sum_{k=1}^{L} \left(\frac{\gamma}{\beta_k}\right) N_k \left(1 - \frac{p_k^i}{\mu_k}\right) + O\left[\left(\frac{p_k^i}{\mu_k}\right)^2 + \left(\frac{p_k^s}{\mu_k}\right)^2\right] \sim S_{total}^0 \qquad (21)$$

Thus, when the rate of movement of susceptible or infected people from one patch to another is small compared to the birthrate of susceptibles, the steady state total number of susceptibles is only moderately different from that total when infected people do not move between the patches.

If there are L patches, the movement of susceptible and infected people between the patches will likely depend on the number of susceptible and infected people already in each patch. Thus the generalization of Eqs. (11-12) will have terms such as $p_k^s = p_k^s(s_1, s_2, s_3, \ldots s_k)$ and $p_k^i = p_k^i(i_1, i_2, i_3, \ldots i_k)$. For the general case, as well as even for the steady state, where $ds_k/dt = di_k/dt = 0$, the equations are nonlinear and there is no simple analytical solution for $s_k$ and $i_k$. However, as we now show, these equations can be solved for the steady state if we make the assumptions that there is no movement of susceptibles and that the rate of movement of infecteds out from a patch is proportional to the number of infecteds in that patch. Therefore, we now use these assumptions to compute the steady state solutions analytically and the transient dynamics numerically.

We now consider the case of two patches with $S_k$ susceptibles, $I_k$ infected, $N_k$ people, contact rate $\beta_{0k} = (\beta_k/N_k)$ and birthrate $\mu_k$, where $k = 1, 2$. We assume that infected people from patch 1 move into patch 2 at a rate of $(r_1\gamma)I_1$ and from patch 2 into patch 1 at a rate of $(r_2\gamma)I_2$. The parameters $r_k$ are the ratios of the rate of decline of the

page 7

number of infecteds in patch k due to those who leave the patch, $r_1(\gamma I_1)$, compared to those recover from the disease, $(\gamma I_1)$. During a brief time $\Delta t$, $(r_1\gamma)I_1\Delta t$ infecteds move from patch 1 into patch 2, and $(r_2\gamma)I_2\Delta t$ infecteds move from patch 2 into patch 1, as illustrated in Fig. 1. For patch 1, the new number of infecteds after a time $\Delta t$ is $[I_1 - (r_1\gamma)I_1\Delta t] + (r_2\gamma)I_2\Delta t$. Thus, the change in $S_1$ and $I_1$ over a brief time $\Delta t$ are

$$\Delta S_1 = [\mu_1 N_1 - \beta_{01}(1-r_1\gamma\Delta t)I_1 S_1 - \beta_{01}(r_2\gamma)I_2 S_1 \Delta t]\, \Delta t \tag{22}$$

$$\Delta I_1 = [\beta_{01}(1-r_1\gamma\Delta t)I_1 S_1 - \gamma(1-r_1\gamma\Delta t)I_1 \\ + \beta_{01}(r_2\gamma)I_2 S_1 \Delta t - \gamma(r_2\gamma)I_2 \Delta t \\ - (r_1\gamma)I_1 + (r_2\gamma)I_2]\, \Delta t \tag{23}$$

In the limit as $\Delta t$ approaches 0, the terms of order $(\Delta t)^2 \ll (\Delta t)$, and Eqs. (22-23) become

$$\frac{dS_1}{dt} = \mu_1 N_1 - \beta_{01} I_1 S_1 \tag{24}$$

$$\frac{dI_1}{dt} = \beta_{01} I_1 S_1 - \gamma I_1 - (r_1\gamma)I_1 + (r_2\gamma)I_2 \tag{25}$$

The full set of equations for the two patches can now be expressed as

$$\frac{ds_1}{dt} = \mu_1 - \beta_1 i_1 s_1 \tag{26}$$

$$\frac{ds_2}{dt} = \mu_2 - \beta_2 i_2 s_2 \tag{27}$$

$$\frac{di_1}{dt} = \beta_1 i_1 s_1 - \gamma i_1 - (r_1\gamma)i_1 + (r_2\gamma)\left(\frac{N_2}{N_1}\right)i_2 \tag{28}$$

$$\frac{di_2}{dt} = \beta_2 i_2 s_2 - \gamma i_2 - (r_2\gamma)i_2 + (r_1\gamma)\left(\frac{N_1}{N_2}\right)i_1 \tag{29}$$

where $s_1 = S_1/N_1$, $s_2 = S_2/N_2$, $i_1 = I_1/N_1$, and $i_2 = I_2/N_2$. The steady state solutions, $ds_k/dt = di_k/dt = 0$, to Eqs. (26-29) are

$$s_1 = \frac{\mu_1}{\beta_1 i_1} \tag{30}$$

$$s_2 = \frac{\mu_2}{\beta_2 i_2} \tag{31}$$

$$i_1 = \frac{(1 + r_2)\left(\frac{\mu_1}{\gamma}\right) + r_2 \left(\frac{N_2}{N_1}\right)\left(\frac{\mu_2}{\gamma}\right)}{1 + r_1 + r_2} \quad (32)$$

$$i_2 = \frac{(1 + r_1)\left(\frac{\mu_2}{\gamma}\right) + r_1 \left(\frac{N_1}{N_2}\right)\left(\frac{\mu_1}{\gamma}\right)}{1 + r_1 + r_2} \quad (33)$$

It is straightforward to extend these results to models with a larger number of patches. For example, we used Maple (Waterloo Maple Inc. 2003) to compute the analytical solution for the steady state for models with three patches.

We studied the dynamical properties of this two patch system by numerical integration. As $s_k$ and $i_k$ may approach zero, to insure better numerical stability, the variables $s_k$ and $i_k$ were first replaced by the logarithmically transformed variables $s'_k = \ln(s_k)$ and $i'_k = \ln(i_k)$ in Eqs. (26-29). (Note that the variables here can possibly have values that represent less than one individual, which could cause the termination of the infection. We will consider these effects due to the disrete nature of the population in future studies.) The equations were then integrated numerically in Matlab using ODE113 with relative tolerance $10^{-6}$ and absolute tolerance $10^{-12}$.

We used this two patch model to determine the response of a new population, $N_2$, introduced into a region of population, $N_1$, with steady state endemic disease. We simulated the introduction of the new population at time $t_0$ over time scale $\tau$ by making $r_k$ time dependent

$$\begin{aligned} r_k(t) &= 0 & t &< t_0 \\ r_k(t) &= r_k^0 [1 - e^{-(t-t_0)/\tau}] & t &> t_0 \end{aligned} \quad (34)$$

A single, isolated patch has oscillations in the fraction of susceptible and infected people about the steady state that define a natural frequency of the patch which can be computed from the eigenvalues of Eqs. (1-2). The period T of this natural frequency is approximately

$$T \sim \frac{2\pi}{\sqrt{\beta\mu}} \quad (35)$$

The dynamical behavior of this two patch model depends strongly on the relationship between $\tau$ and T. We illustrate this with $N_1 = 10^6$, $N_2 = 5 \times 10^5$, $r^0_1 = 0.1$, $r^0_2 = 0$, $t_0 = 10$ yr and the parameters typical of those of measles [6] namely, $\gamma = 100$ yr$^{-1}$, $\mu_1 = 0.02$ yr$^{-1}$, and $\beta_1 = 1200$ yr$^{-1}$. Based on the scaling assumption of Eqs. (3-4), we chose $\mu_2 = 0.02$ yr$^{-1}$ and $\beta_2 = 1000$ yr$^{-1}$, namely that $\varepsilon = 0$, and $\delta = -0.263$. For these parameters, the natural period of the patches from Eq. (35) are $T_1 = 1.3$ yr and $T_2 = 1.4$ yr.

For $\tau = 10$ yr $\gg T_k$, the onset time scale of the migration of infected people is much longer than the natural time scale of the patches. In this case the migration does not induce a significant epidemic in either population. This is illustrated in Figure 2a and 2b. Also shown in Figure 2 is the adiabatic, quasi-steady approximation for $s_1$, $s_2$, $i_1$, and $i_2$ as defined by evaluating the steady state solutions, Eqs. (30-33), using the time dependent values of $r_k(t)$ from Eq. (34).

However, for $\tau = 0.1$ yr $\ll T_k$, the onset time scale of the migration of infected people is much shorter than the natural time scale of the patches. Now the migration induces significant epidemics (local maxima in $i_k$) in both populations. This is illustrated in Figure 3. Note that the largest epidemic in the newly introduced population is not directly driven by the initial flux of infected people into it from the large regional population. The sudden loss of infected people from the larger, regional population eventually drives it far from its own steady state, inducing epidemics in its own population. The flux of infecteds into the newly introduced population is proportional to the number of infecteds in the large, regional population. Therefore it is these epidemics in the large regional population that now induce epidemics in the newly introduced population. It is also striking that quite small values of $r_k$ can drive significant epidemics, when $\tau \ll T_k$. The sensitivity of patch dynamics to such small coupling parameters between the patches has been noted by Viz, Billings, and Schwartz [18] in models where the patches are coupled by mass action terms. At long times, $s_k$ and $i_k$ converge to the same steady state solutions as in Figure 2.

The strength of the largest migration induced epidemic in the newly introduced population, as measured by max($i_2$) from the numerical simulations, is illustrated in Figure 4 as a function of ($\tau/T_1$), for $r_2 = 0$ and several values of $r_1$. When ($\tau/T_1$) $\gg 1$, the initial local maximum in $i_2$ is lower than the new steady state value of $i_2$ reached at long times, and thus max($i_2$) is equal to this new steady state value of $i_2$. As ($\tau/T_1$) decreases, the maximum of the migration induced epidemics increases and reaches a plateau as ($\tau/T_1$) $\ll 1$. The plateau occurs because the epidemics in the larger regional population (and the subsequent epidemics it induces in the newly introduced population) depend on the natural period, $T_1$, of that patch and therefore become insensitive to further decreases in $\tau$. As expected, the transition between the new steady state values and the plateau is reached at approximately ($\tau/T_1$) $\sim 1$.

We also studied the effects of turning off the migration of infected people between the patches. When the patches were at their steady state values at constant values of migration, we reduced the migration to zero over an offset time scale of $\tau$. Similar to results we found for the onset of migration, when the offset times scale $\tau \gg T_k$ there were no migration induced epidemics while $\tau \ll T_k$ initiated migration induced epidemics.

An important conclusion from these two patch flux-based studies is that the rate of migration of infected people, $r_k$, only moderately alters the steady state fraction of infected people in each patch, as given by Eqs. (32-33), and it alone does not determine if there are migration induced epidemics. It is the onset time scale, $\tau$, of this migration that determines the existence and severity of epidemics in both populations. Migration induced epidemics occur when the onset time scale is less than the natural time scale T of the patches, which is given by Eq. (35). Moreover, the severity of the migration induced epidemics reaches a plateau and does not increase further when the onset time scale is further reduced beyond approximately 0.1 that of the natural time scale of the patches.

In order to model more than two patches we extend equations (26-29) to k = 1, 2, ... L patches by defining an L x L matrix r, where $r_{kj}\gamma I_k$ is the flux of infected from patch k to patch j. The equations, with the logarithmically transformed variables $i'_k = \ln(i_k)$ and $s'_k = \ln(s_k)$, are now

$$\frac{ds'_k}{dt} = \mu_k e^{-s'_k} - \beta_k e^{i'_k} \qquad (36)$$

$$\frac{di'_k}{dt} = \beta_k e^{i'_k} - \gamma + \gamma \sum_{j=1}^{L} [r_{jk}(\frac{N_j}{N_k})e^{i'_j - i'_k} - r_{kj}] \qquad (37)$$

We studied several different patterns of migration between the patches including: equal migration between all the patches, randomly assigned values for the migration between all the patches, equal serial migration (patch 1 <--> 2 <--> 3 ... <--> L), equal unidirectional migration (patch 1 --> 2 --> 3 ... --> L), equal ring migration (patch 1 <--> 2 <--> 3 ... <--> L <--> 1), as well as uncoupled patches. We set fixed ratios, $N_k/N_{k+1}$, for the populations between the patches and used the scaling assumptions of Eqs. (3-4) to choose the values of $\mu_k$ and $\beta_k$ in the patches. We computed the results for models with L ≤ 20.

As the degree of coupling is increased, for example, from equal serial migration to equal migration between all the patches, the epidemics in all the patches become more synchronous and more coupled to the natural time period of the patch with the largest population, even at very low values of r. This is consistent with results from other studies of coupled systems where the coupling strength to produce global synchrony approaches zero as the system becomes globally connected [19,20].

Another interesting result from these studies is illustrated in Figures 5 and 6. In these simulations 20 patches are serially connected: $r^0_{k,k+1} = 0.002$ (k = 1, 2, ... 19), the parameters of the first patch are the same as those in the two patch model, the populations of the subsequent patches are $N_k/N_{k+1} = 2$, $\mu_k/\mu_{k+1} = 1$ corresponding to the scaling parameter $\epsilon = 0$, $t_0 = 10$ yr, and $\tau = 0.1$ yr. When $\tau \ll T_k$ and an epidemic is induced by a rapid onset of migration, the patch with the largest population drives epidemics in the other patches. Each patch has its own natural period given by Eq. (35) and so responds at its own time scale and therefore reaches its local maximum of the fraction infected at that time scale. When the natural period of the patches is tuned to increase along the line of patches away from the first largest patch, then the patches further from the first, largest patch reach their maximum more slowly. It appears as if a wave of epidemics sweeps away from the first, largest patch. This is illustrated in Figure 5, where $\beta_k/\beta_{k+1} = 1.1$, which corresponds to a scaling parameter of $\delta = -0.137$. On the other hand, when the natural period of the patches is tuned to decrease along the line of patches away from the first largest patch, then the patches further from the first, largest patch reach their maximum faster. It appears as if a wave of epidemics sweeps towards the first, largest patch. This is illustrated in Figure 6, where $\beta_k/\beta_{k+1} = 0.9$, which corresponds to a scaling parameter of $\delta = 0.152$.

    Traveling waves have been observed, for example, in the spread of dengue haemorrhagic fever in Thailand [21]. Such spatial traveling waves of epidemics can be generated by spatial transmission through diffusion or stochastic fade-out and reintroduction of disease in adjacent patches [2, 21]. However, the apparent wave of epidemics that travels along the line of serially connected patches presented here is a different phenomena. These patches form a serial line of (nonlinear) oscillators, each with their own time scale given by Eq. (35). They each reach their maximum response at their individual time scale in response to the sudden drive exerted simultaneously in all the patches by the rapid onset of the migration. The apparent wave here arises from the different time delays for each individual patch to reach its maximal response. This contrasts with a true traveling wave advancing by spatial transmission through consecutive patches. We do not argue that the phenomena found here is the cause of the observed traveling waves in the experimental data. However, these simulations do demonstrate that such a phenomena is possible when there is a variation of infectious parameters $\beta$ and $\mu$ across a region, which may reasonably arise from the variation in population densities, social factors, or geographic factors across a region. Therefore, our results do suggest that such a mechanism should at least be considered when an apparent traveling wave of epidemics is observed across regions with spatial gradients of infectious parameters.

    In summary, classical models of the spread of infectious diseases have assumed that the infected and susceptible people are either well mixed or that separate patches of people interact through a mass action term. Here, we have derived the equations that model the spread of infection driven by the actual migration of people between the patches. We determined the steady state solutions for multiple patches with power law scalings of the infectious parameters and compared those results to that of a well-mixed population. We showed, perhaps surprisingly, that the steady state total number of infected people is the same whether or not susceptible and infected people move between the patches. Then we studied the dynamical response when the flux of infecteds between the patches is turned on at different rates. This flux-based patch approach is useful in modeling the transient response when a new population is moved into a region

with endemic disease. Epidemics result when the time scale of the onset of this migration is rapid compared to the natural time scale of the patches. These migration induced epidemics can take on the appearance of traveling waves across serially connected patches. For example, for the parameters used in the model illustrated in Figure 5, the wave travels from the largest patch to the furthest patch in approximately 6 months, of the same order as that reported for the traveling wave in the spread of dengue haemorrhagic fever in Thailand [21]. Therefore, if there is a spatial gradient of infectious parameters across that region, such a mechanism should be considered as one possible explanation for this wave. The results from these models demonstrate the importance of determining the spatial distribution of the infectious parameters from experimental data because of the role they play in the dynamics of the spread of disease.

**Acknowledgments**

This work was supported by the Office of Naval Research and the Center for Army Analysis. LSL was supported by the 2004 ONR-ASEE Summer Faculty Research Program at the Naval Research Laboratory.

**Figure Legends**

Figure 1. Flux-based model for the migration of infected people between two patches. During each brief time $\Delta t$, $(r_1\gamma)I_1\Delta t$ infected people move from patch 1 into patch 2, and $(r_2\gamma)I_2\Delta t$ infected people move from patch 2 into patch 1.

Figure 2. Fraction of susceptible (a) and infected (b) people in each of two patches computed from the two patch flux-based model. At $t_0 = 10$ yr infected people from the large regional population, $N_1 = 10^6$, start migrating into the newly introduced population $N_2 = 10^5$, at onset time scale $\tau = 10$ yr. At short times, there is no significant epidemic in either patch since the onset time scale of the migration is much longer than the natural time scale of the patches. At long times, the migration changes the steady state fraction of susceptible and infected people. Legend: bold black line = large regional population, light black line = newly introduced population, dotted lines = adiabatic quasi-steady state approximations, boxes = maxima. (The dotted line for the adiabatic quasi-steady state approximation is covered by the overlapping bold black line for the susceptibles in the large regional population in Figure 2a).

Figure 3. Same two patch flux-based model as Figure 2, except that the onset time scale of the migration of infected people, $\tau = 0.1$ yr is now much shorter than the natural time scale of the patches. There are now significant epidemics (local maxima in $i_k$) in both populations. Note that the largest epidemics in the newly introduced population are driven by the epidemics in the large regional population that were triggered by the initial flux of infected people out of this population. At long times, $s_k$ and $i_k$ converge to the same steady state solutions as in Figure 2.

Figure 4. The strength of the migration induced epidemic in the newly introduced population (max $i_2$) is plotted as a function of the ratio of the onset time scale of the migration of infected people to the natural frequency, namely, $\tau/T_1$, for $r_1 = r$ and $r_2 = 0$. As $\tau/T_1$ decreases, the maximum fraction of infected people in the newly introduced population increases and reaches a plateau.

Figure 5. Flux based model with 20 patches (k = 1, 2, 3, ... 20) serially connected in a line. The ratio of $\beta_k/\beta_{k+1} = 1.1$, so that the natural period increases away from the first, largest patch (k = 1) along the line until the last patch (k = 20). The peaks (boxes) of the fraction of infected people in the migration induced epidemic appear to travel as a wave away from the first patch.

Figure 6.  Flux based model with 20 patches (k = 1, 2, 3, ... 20) serially connected in a line. The ratio of $\beta_k/\beta_{k+1} = 0.9$, so that the natural period decreases away from the first, largest patch (k = 1) along the line until the last patch (k = 20).  The peaks (boxes) of the fraction of infected people in the migration induced epidemic appear to travel as a wave towards the first patch.

**Figure 1**

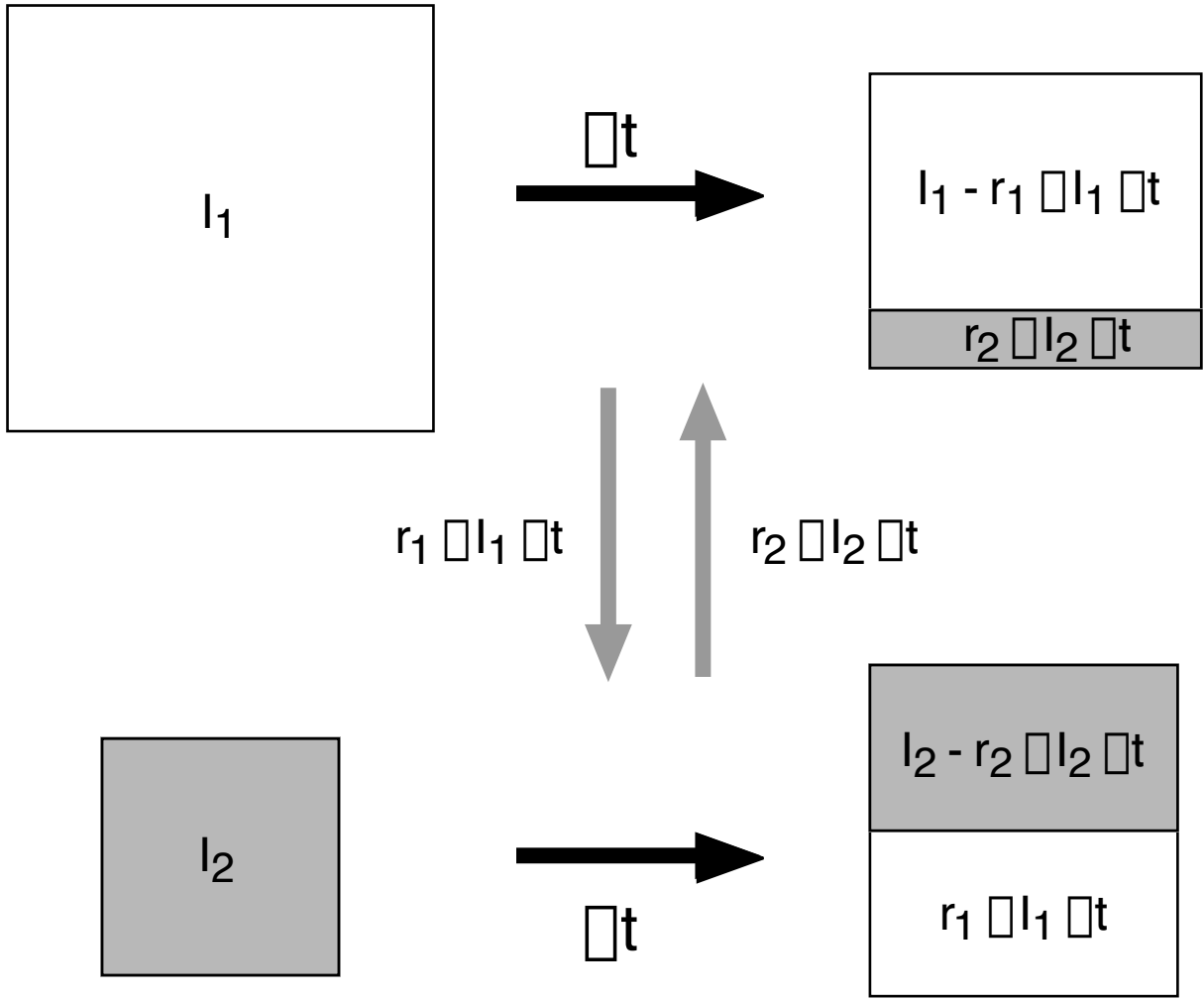

**Figure 2a**

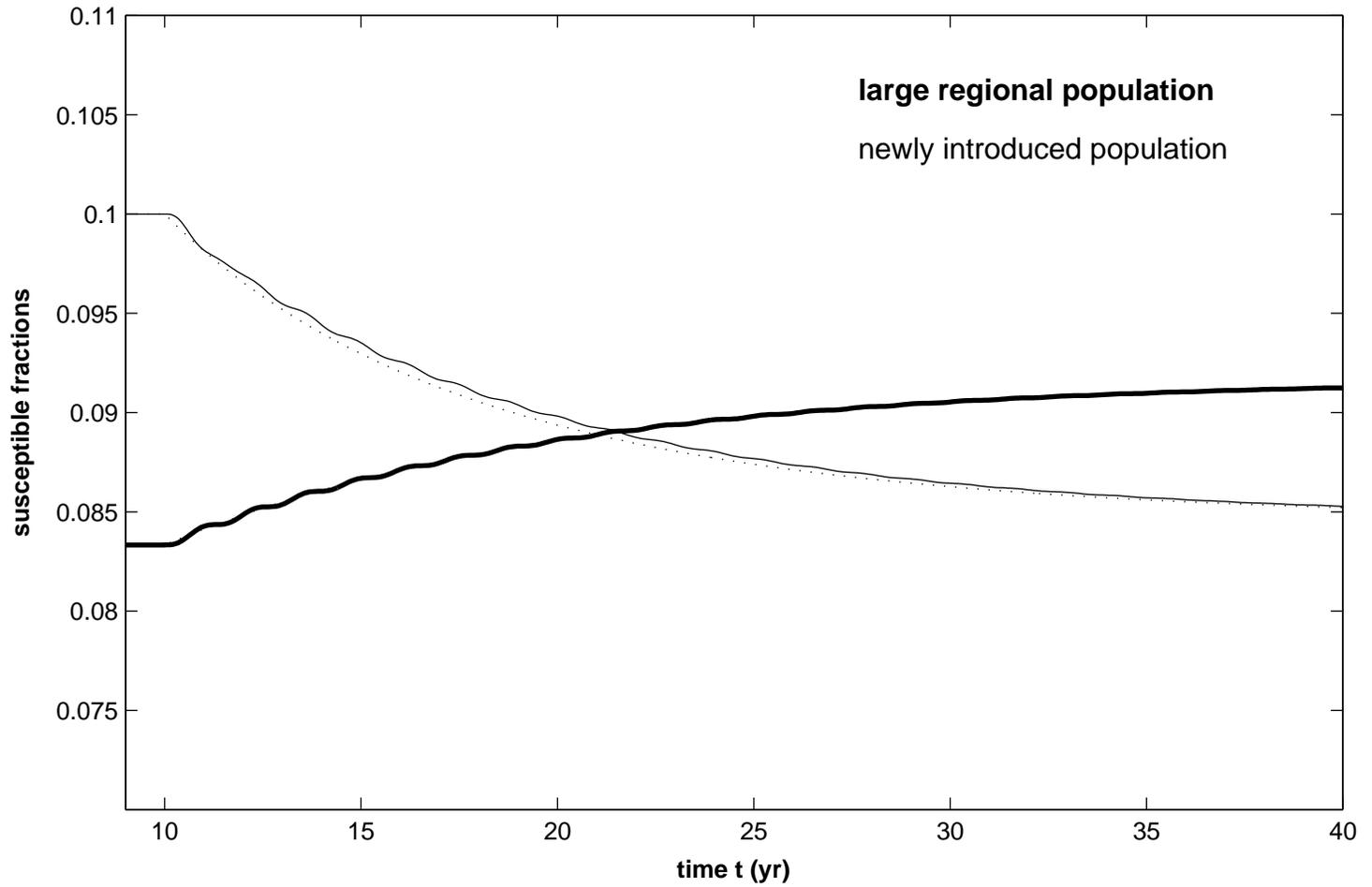

**Figure 2b**

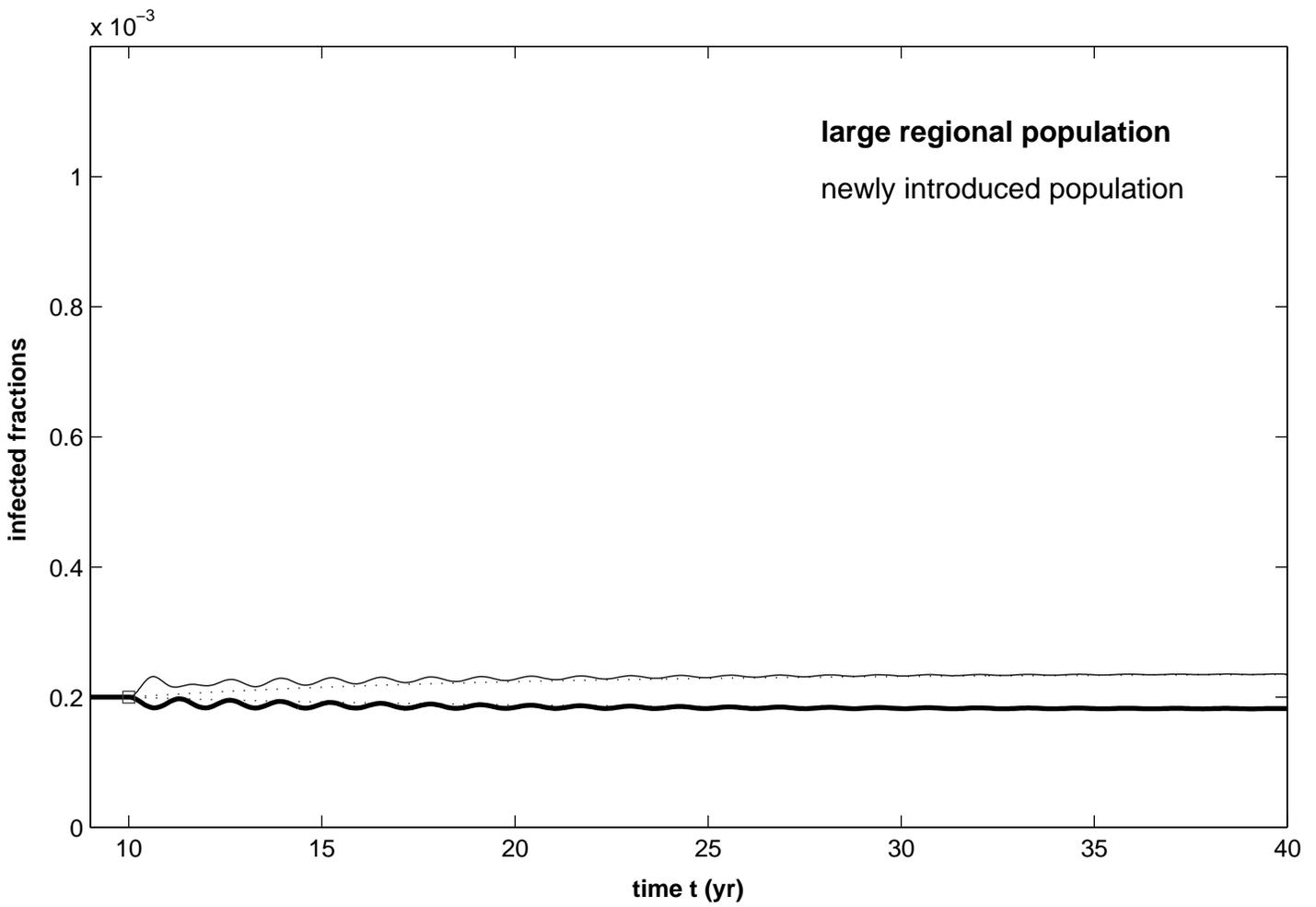

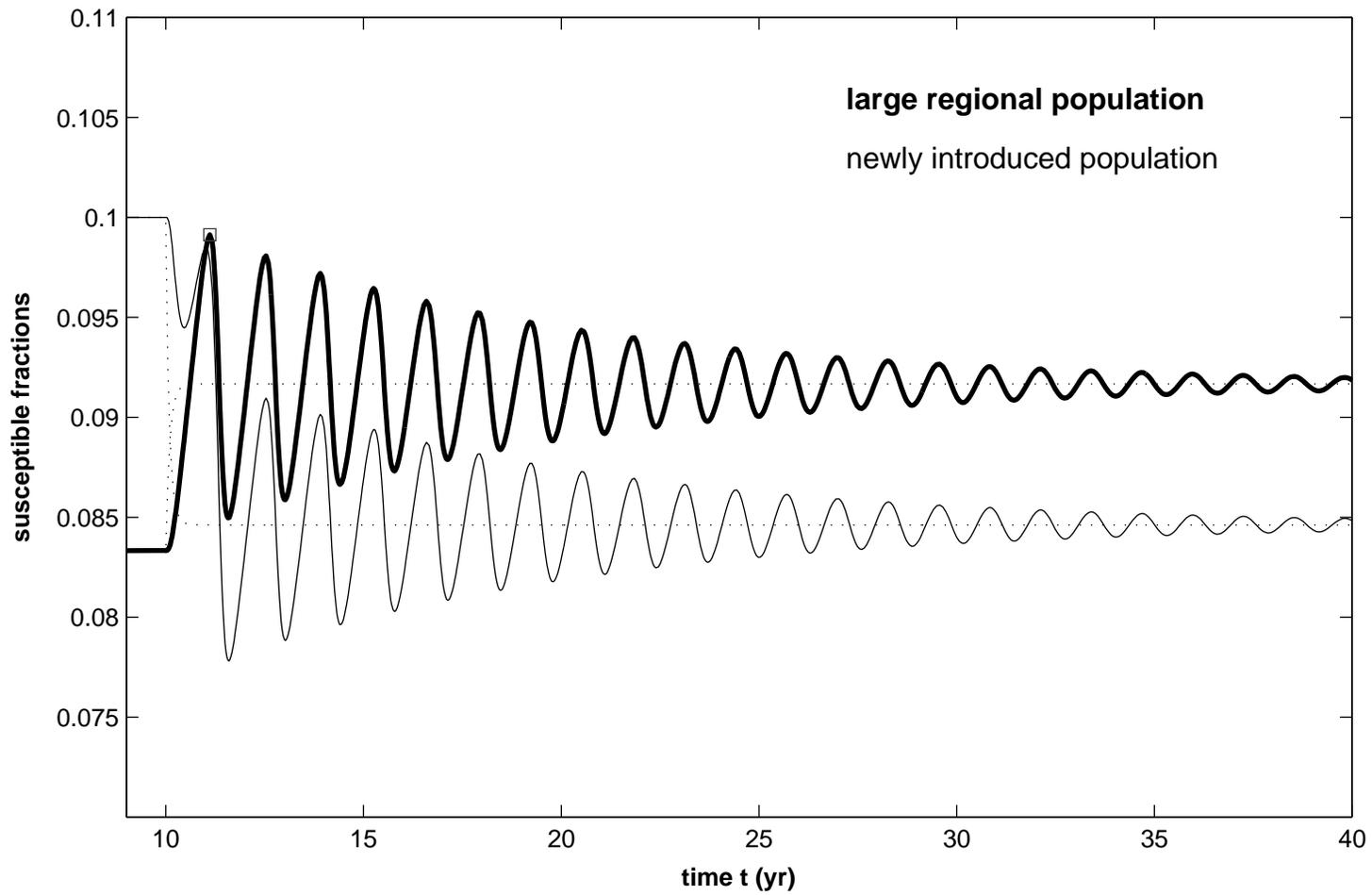

**Figure 3a**

**Figure 3b**

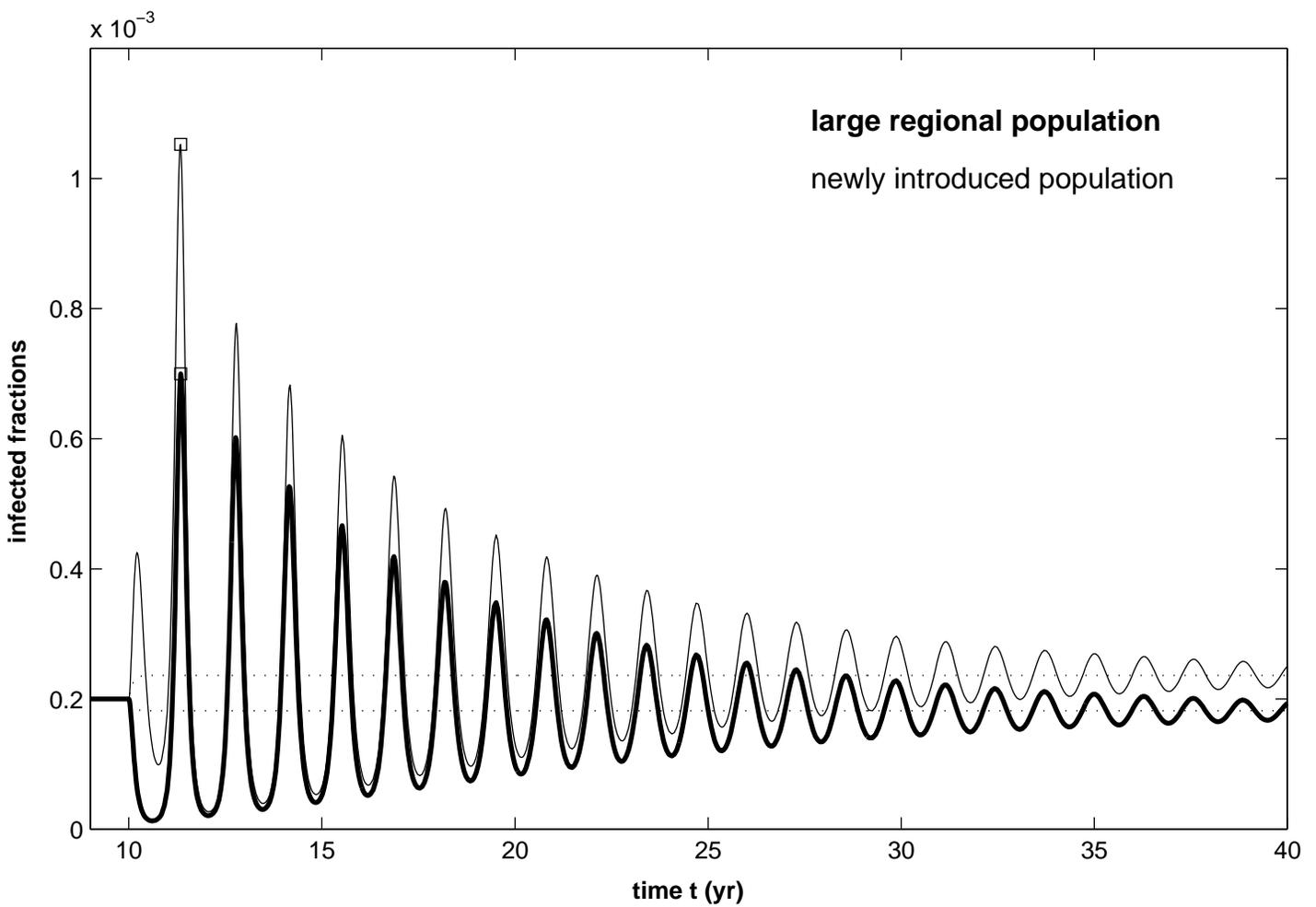

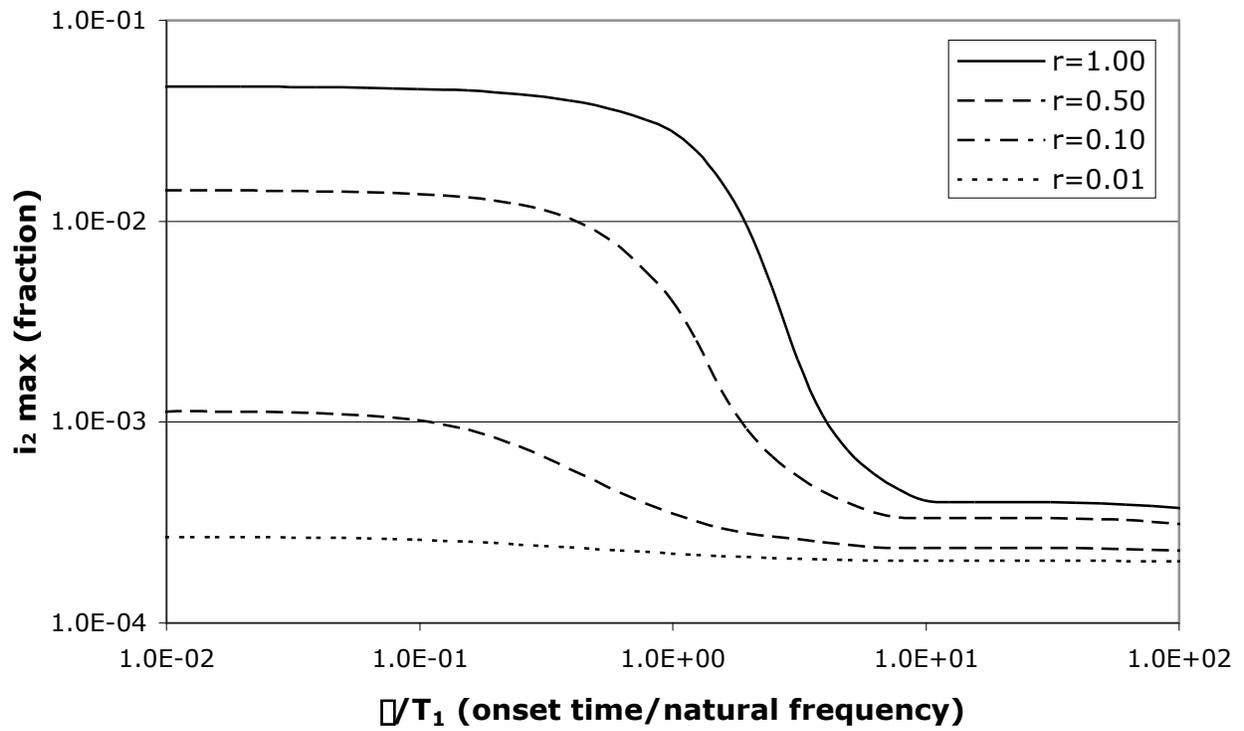

**Figure 5**

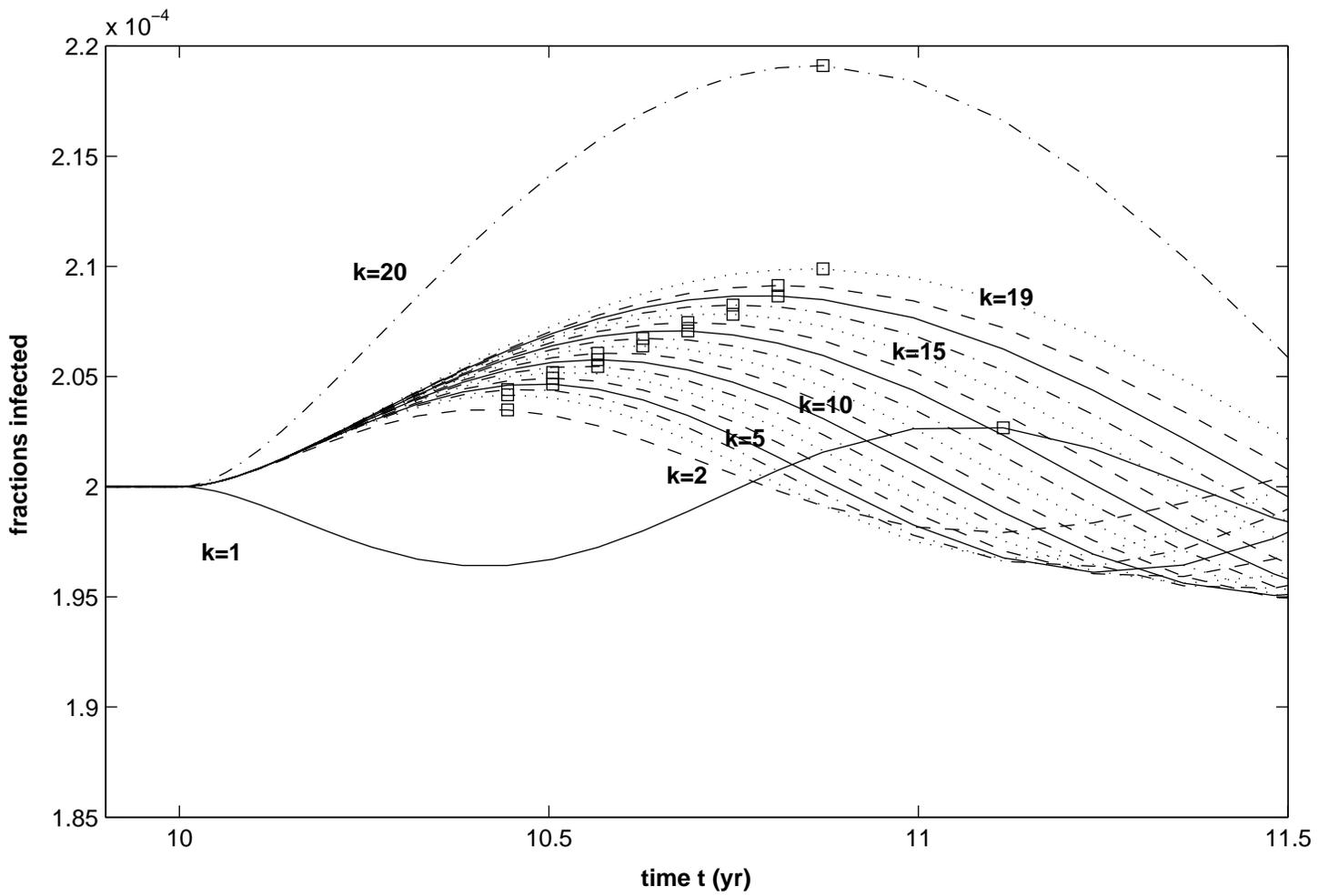

**Figure 6**

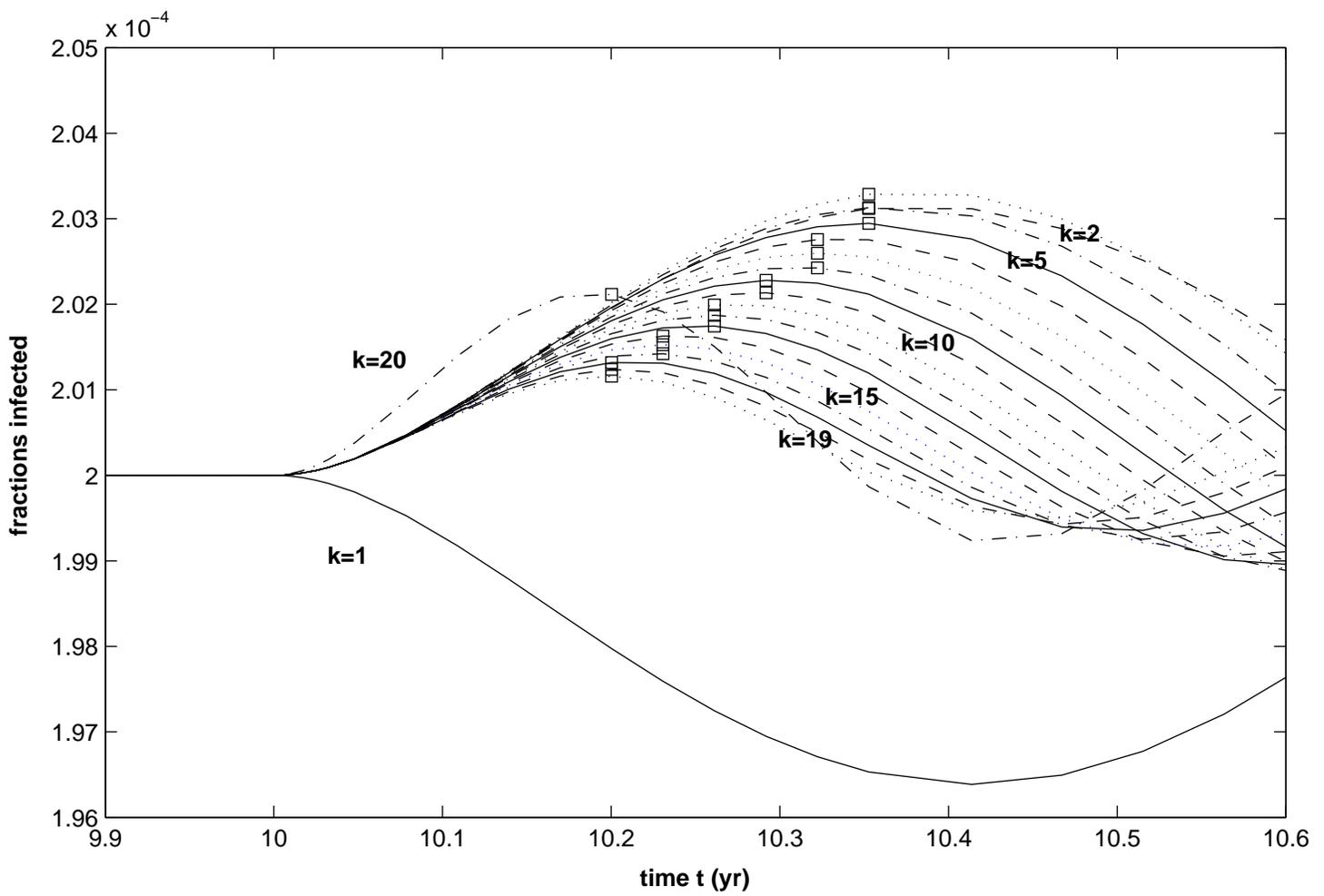